# Majorana Edge-to-bound State Transition in Interferometric Superconductor Rings

Xiangyu Bi[1,2†], Aoshen Yang[1,2†], Hongzi Cao[1,2†], Ganyu Chen[1,2†], Yuqiang Fang[3†], Zeya Li[4], Qizheng Sun[1,2], Junwei Huang[1,2], Da Wang[1,5], Fuqiang Huang[3*], Toshiya Ideue[6], Qianghua Wang[1,5,7*], Yoshihiro Iwasa[8*], Hongtao Yuan[1,2,7*]

[1] *National Laboratory of Solid State Microstructures and Collaborative Innovation Center of Advanced Microstructures, Nanjing University, Nanjing 210093, China.*

[2] *College of Engineering and Applied Sciences, Nanjing University, Nanjing 210023, China.*

[3] *Key laboratory of Intelligent Creation for Extreme Energy Materials of Ministry of Education, School of Materials Science and Engineering, Zhangjiang Institute for Advanced Study, Shanghai Jiao Tong University, Shanghai 200240, China.*

[4] *Quantum Science Center of Guangdong-Hong Kong-Macao Greater Bay Area, Shenzhen 518045, China.*

[5] *School of Physics, Nanjing University, Nanjing 210093, China.*

[6] *Institute for Solid State Physics, The University of Tokyo, Tokyo 113-8654, Japan.*

[7] *Jiangsu Physical Science Research Center, Nanjing 210093, China.*

[8] *RIKEN Center for Emergent Matter Science, Wako 351-0198, Japan.*

[†]These authors contributed equally to this work.

*Correspondence to: htyuan@nju.edu.cn (H.T.Y.), iwasay@riken.jp (Y.I.), qhwang@nju.edu.cn (Q.H.W.), huangfq@sjtu.edu.cn (F.Q.H.).

**Manipulating flux quanta in topological superconductors is central to vortex braiding in quantum computing. Their interferometric superconducting rings can host either topologically protected edge states with fluxes confined inside the central hollow area or vortex bound states in flux cores at the annulus walls. Here we demonstrate a strategy to drive topological edge states into vortex bound states by tuning the relative scale of vortex size and annulus width in nano-ring interferometric device on 2M-$WS_2$ topological superconductor. Near its superconducting transition where the coherence length exceeds the annulus width, Little–Parks oscillations exhibit expected periods for flux bundles in ring-center, consistent with edge states residing at inner and outer rims of the annulus. Upon cooling, where the coherence length shrinks below the annulus width, the unexpectedly increased oscillation period reflects multiple fluxes distributed along annulus region in flux necklace configuration, implying a transition from extended edge states to localized vortex bound states. Ginzburg–Landau simulations and local density-of-states calculations with Bogoliubov-de Gennes equations reveal that large values for both Ginzburg–Landau parameter and flux-pinning energy are vital for such a transition. Our findings provide a novel strategy for understanding flux manipulation towards flux braiding and topological quantum computing.**

Flux quanta and their spatial distribution in superconductors play key roles in understanding flux dynamics in condensed matter physics and superconducting electronics[1–5]. When applying a perpendicular magnetic field to a superconducting nano-ring system, the magnetic flux quanta $\Phi_0 = h/2e$ ($h$ is the Planck's constant and $e$ is the elementary charge) can enter its central hole[6,7], and the resulting flux distribution is typically governed by the relation between the annulus width $w$ of the nano-ring and the in-plane superconducting coherence length $\xi_{ab}$. When $w < \xi_{ab}$, magnetic flux quanta can penetrate the central hole of the nano-ring, forming a "flux bundle" configuration (Fig. 1a). Such a configuration has been experimentally demonstrated in β-$Bi_2Pd$ nano-rings[8], chiral $TaS_2$ superlattices[9] and nanopatterned cuprates[10], with the Little–Parks oscillation as a sensitive probe for the flux quantization and phase coherence[11]. In sharp contrast, in the regime $w > \xi_{ab}$, fluxes can penetrate through the annulus wall of the superconducting nano-ring and can thus induce a series of flux quanta along the annulus wall, forming a chain-like "flux necklace" configuration instead (Fig. 1b), which can in principle induce an extremely large Little–Parks oscillation period. In topological superconducting interference nano-rings, theoretical rationale is that the flux quanta can give rise to either topologically protected edge states forming at the inner/outer rim of the annulus in a flux-bundle configuration, or vortex bound states localized in a flux-necklace configuration that might host Majorana zero modes in their vortex cores[12–14]. However, experimentally realizing such a flux-configuration-driven topological modes transition remain challenging and elusive.

Here, we demonstrate a universal, coherence-length-driven strategy to drive the flux configuration transition from a central flux bundle to a necklace-like flux distribution along the annulus wall in interferometric nano-rings of topological superconductor 2M-$WS_2$. Near the superconducting transition regime with the coherence length larger than the annulus width, the observed Little–Parks effect with expected oscillation periods corresponds to the flux-bundle formed within the hollow ring center, where the topologically protected edge states distribute at inner and outer rims of the annulus. In contrast, by cooling the temperature to the level with coherence length smaller than the annulus width, the greatly enlarged Little–Parks oscillation period corresponds to multiple fluxes distributing along the annulus superconducting region, implying the emergence of localized vortex bound states that are predicted to harbor Majorana zero modes. Being a sensitive tool to probe the flux distribution in a superconducting nano-ring structure, magnetoresistance curves with great enlargement of oscillation period can help us experimentally verify this coherence-length-driven flux configuration transition. In local density-of-states calculations based on Bogoliubov–de Gennes equations, cusp-like zero-energy peaks appear at the vortex cores in the flux necklace configuration, supporting the theoretical expectation of Majorana bound states therein, and the topological edge-to-bound state evolution is also well captured. Our time-dependent Ginzburg–Landau (TDGL) simulations of flux quanta distribution reveal that large values for both Ginzburg–Landau parameter and flux-pinning energy are vital for stabilizing such a flux necklace configuration, confirmed by our controlled experiments with Nb and 2H-$NbSe_2$ nano-rings where only the conventional Little–Parks oscillation with the flux bundle configuration can be observed. Our observations of flux configuration transition in interferometric superconducting nano-rings establish a novel strategy for designing flux-interference devices that exploit configurable topological states towards vortex braiding and topological quantum computing.

**Interferometric superconducting nano-rings with unconventional Little–Parks oscillations**

We chose topological superconductor candidate 2M-$WS_2$ as the target material for interferometric nano-ring device to probe the flux configuration transition because of the following three reasons. First, 2M-$WS_2$ has been confirmed hosting the topological superconductivity induced by the intrinsic proximity effect between bulky *s*-wave superconducting state and topological surface state[15–18], where the Majorana bound states located at the vortex cores have been directly confirmed by the zero-bias conductance peak in d*I*/d*V* spectra[15]. The high transition temperature $T_c \approx 8.8$ K, together with its topological superconducting nature, enables the penetration of flux quanta carrying Majorana bound states in their vortex cores[15], and allows us to investigate the flux distribution within such a nano-ring structure. Second, the charge stripe phase in 2M-$WS_2$ nanoflakes[17,19] can naturally serve as local energy wells to pin the flux quanta, thus stabilizing the flux distribution and making it energetically preferential for flux quanta arranging along the annulus wall to form the flux necklace configuration (Fig. 1a,b) for experimental observation. Third, the wide paraconductivity regime of superconducting transition in electronic transport measurements in thin 2M-$WS_2$ nanoflakes[16,17] (details in Supplementary Figs. 1-3) provides a broad temperature range to understand flux configuration evolution with different $\xi_{ab}$ values, and to capture the flux configuration transition with Little–Parks oscillations.

To determine the $\xi_{ab}$ value and its temperature evolution, we measured temperature ($T$) dependence of four-terminal resistance ($R_{xx}$) at different out-of-plane magnetic fields (Fig. 1c). Figure 1d shows the evolution of upper critical magnetic field and coherence length with temperature. One can see that the $\xi_{ab}$ value varies from ~70 nm (8.2 K) to 15 nm (2.0 K), where the out-of-plane upper critical magnetic field $\mu_0 H_{c2}^{\perp}$ (blue balls) can be defined by 50% of normal-state magnetoresistance at 10 K, and the in-plane coherence length $\xi_{ab}$ is obtained with Ginzburg-Landau theory based on $\mu_0 H_{c2}^{\perp} = \Phi_0/2\pi\xi_{ab}^2$ equation[20]. The critical condition for the flux necklace configuration is set to be $\xi_{ab} \approx w/5$ based on the standard for the formation of Pearl vortices within the annulus wall[21,22]. Therefore, we designed the width of annulus wall $w$ to be approximately 200 nm for 2M-$WS_2$ nano-ring (inset of Fig. 1d), and thus we will have ~40 nm $\xi_{ab}$ value as the critical condition to form the flux necklace configuration in 2M-$WS_2$ nano-rings (black dashed line in Fig. 1d). Note that such a designed geometry can provide us with a suitable critical condition of $\xi_{ab}$ which crosses over the experimentally obtained coherence length range, and thus simply by tuning the temperature can we experimentally switch the flux configuration between the bundle and necklace states.

To understand the flux distribution in 2M-$WS_2$ superconducting nano-rings, we performed Little–Parks oscillation measurements and compared their oscillation period with theoretical expectations. Figure 1e shows the Little–Parks oscillation of a 2M-$WS_2$ nano-ring with the thickness of 27 nm, where the upper panel shows the original magnetoresistance curve and the bottom panel shows the pure oscillation signal $\Delta R_{xx}$ by subtracting the background signal via Savitzky-Golay filter[23,24]. In the fast Fourier transform of pure oscillation data, one can see a cusp peak located at ~220 $T^{-1}$ (inset of Fig. 1e), which corresponds to an experimentally-confirmed oscillation period of ~4.5 mT. Theoretically, the oscillation period $\mu_0 \Delta H$ can be estimated based on the equation $\mu_0 \Delta H = \Phi_0/\pi r_{\mathrm{eff}}^2$, where the effective radius of the interference path for supercurrent $r_{\mathrm{eff}} = \sqrt{\frac{r_{\mathrm{in}}^2 + r_{\mathrm{out}}^2}{2}}$ is obtained from the root mean square of inner ($r_{\mathrm{in}}$) and outer ($r_{\mathrm{out}}$) radius of the nano-ring[25]. The calculated conventional Little–Parks oscillation period is approximately 4.49 mT based on the $r_{\mathrm{in}}$ and $r_{\mathrm{out}}$ values of ~255 and

~478 nm (obtained from the scanning electron microscopy image in Fig. 1d), and in excellent agreement with the experimentally-confirmed period, indicating the formation of conventional flux bundle configuration therein. Note that the phase-angle of the oscillation for such a flux bundle scenario is 0 (zero field corresponds to the local minima of $\Delta R_{xx}$) rather than π, in accordance with the *s*-wave pairing nature of 2M-$WS_2$ nanoflakes[17]. Temperature-dependent mapping results of $\Delta R_{xx}$ (Fig. 1f) show that the oscillation period remains unchanged as the temperature decreases, indicating that the system is stabilized in the flux bundle configuration with Aharonov-Bohm type magnetic fluxes[26] located at the hollow center of nano-ring.

Interestingly, unconventional Little–Parks oscillation appears for a 5-nm-thick 2M-$WS_2$ nano-ring device with large superconducting paraconductivity transition regime. Figure 2a shows the temperature dependence of $R_{xx}$ for such a thin 2M-$WS_2$ nano-ring, where a wide temperature range of superconducting transition can be confirmed based on its first-order derivative $dR_{xx}/dT$ (Supplementary Fig. 4). For the higher temperature above 6.8 K (light blue area, corresponds to a larger $\xi_{ab}$ value), the observed Little–Parks oscillation period (blue shaded area in Fig. 2b) also follows the relation $\mu_0 \Delta H = \Phi_0/\pi r_{\mathrm{eff}}^2$ discussed above (Fig. 1e), indicating the formation of flux bundle configuration at the hollow center of the nano-ring. In sharp contrast, when the temperature decreases below 4.4 K (light red area, corresponds to a smaller $\xi_{ab}$ value), a unique Little–Parks oscillation with largely increased period can be observed (denoted as unconventional Little–Parks oscillation, red shaded area in Fig. 2b), implying the formation of flux necklace configuration rather than the flux bundle (discussed later)[27]. However, for the temperature range from 6.8 K to 4.4 K (light yellow area), no apparent magnetoresistance oscillations can be observed since the $\xi_{ab}$ value is comparable to $w$ in such a case.

To compare the difference between conventional and unconventional Little–Parks oscillation phenomena and their relations to flux configurations, we analyzed the oscillation amplitude and period for both scenarios. Figure 2c presents the temperature dependence of oscillation amplitude for both conventional (blue curve) and unconventional (red curve) Little–Parks oscillations. Note that the amplitude of Little–Parks oscillation can be written as[10]

$$R_{\mathrm{osc}} = 0.14T_{\mathrm{c}}\left(\frac{\xi_{ab}\,(0)}{r_{\mathrm{eff}}}\right)^2\frac{\mathrm{d}R_{xx}}{\mathrm{d}T}, \tag{1}$$

where the $\frac{\mathrm{d}R_{xx}}{\mathrm{d}T}$ value can be determined based on the lower panel of Fig. 2a, and $\xi_{ab}\,(0)$ is the extrapolated in-plane coherence length at 0 K, which can be determined via the relation[20]

$$\xi_{ab}(T) = \frac{\xi_{ab}(0)}{\sqrt{1-\frac{T}{T_{\mathrm{c}}}}} \tag{2}$$

Based on the fitting results in Fig. 1d, the $\xi_{ab}\,(0)$ value can be estimated as ~11 nm. Therefore, together with the maximum $\frac{\mathrm{d}R_{xx}}{\mathrm{d}T}$ value obtained from Fig. 2a and the calculated effective radius of interference paths from the oscillation period, the maximum of oscillation amplitude for conventional (or unconventional) Little–Parks oscillation can be theoretically estimated to be ~0.26 Ω (~0.24 Ω) based on Eq. (1). If we compare the amplitude of conventional Little–Parks oscillation (blue curve in Fig. 2c, maximum 0.23 Ω), such an experimental amplitude value is consistent with the theoretical expectation of 0.26 Ω, which agrees with the period comparison for Little–Parks oscillation. However, in unconventional Little–Parks oscillation, the observed amplitude (red curve in Fig. 2c, maximum 0.57 Ω) is much larger than theoretical expectation of 0.24 Ω. Such an unexpected large value for oscillation magnitude might originate from thermally activated[10] or electrical-current-activated[28] vortex motion mechanisms. Since the observed unconventional oscillation shows the periodic behavior, the thermally activated vortex motion scenario might better explain our observations.

Figure 2d,e presents the temperature-dependent mapping of oscillation signal for conventional and unconventional Little–Parks oscillations. Compared to the conventional oscillation period of ~5 mT under expectation (Fig. 2d), the Little–Parks oscillation period is as large as ~70 mT (Fig. 2e), and the corresponding $r_{eff}$ value of 97 nm is much smaller than that of conventional oscillation (~360 nm), which indicates that the conventional flux bundle configuration cannot be used as the model to explain the observed unconventional oscillation period while the flux necklace configuration with the effective oscillation path locating in the annulus wall can explain such phenomena. Note that the calculated effective size ($r_{eff} \approx 97$ nm) of interference path is also comparable with the sizes of fluxes located in charge stripe phase[17], implying that the charge stripe phases at the annulus wall of the 2M-$WS_2$ nano-ring might serve as the pinning center to stabilize the flux distribution. The above-mentioned phenomena are highly reproducible among all the nano-ring devices (details in Supplementary Figs. 5-9 and Supplementary Table 1,2). To further prove that the observed large oscillation period is independent of the formation of a giant vortex state, we performed controlled experiment based on a superconducting nano-disk device with the same $r_{out}$ value as that of the nano-ring device (details in Supplementary Fig. 10). One can see that the above oscillation phenomena were absent in the magnetoresistance curves of the $WS_2$-based nano-disk devices, which excludes the hypothesis that giant vortex state is formed in such a superconducting nano-ring.

To accurately calculate the oscillation periods of the observed Little–Parks oscillations, we performed fast Fourier transform analysis for the pure oscillation signal for both types of oscillations, and the positions of the cusp peaks can be confirmed as 208 $T^{-1}$ (conventional oscillation, Fig. 2f) and 15 $T^{-1}$ (unconventional oscillation, Fig. 2g). For conventional oscillation scenario, the oscillation period and effective radius $r_{\mathrm{eff}}$ of interference path can be confirmed to be 4.8 mT and ~370 nm (illustrated by red dashed circle from the inset of Fig. 2f), and in good accordance with the scenario of flux bundle configuration. However, for the unconventional Little–Parks oscillation, the calculated oscillation period is ~67 mT. Assuming that the interference path is still a circular shape at the hollow center, the effective radius $r_{\mathrm{cal}}$ of interference path can be deduced as ~100 nm. Such a value (red dashed circle in the inset of Fig. 2g) is even far smaller than the inner radius of the nano-ring, indicating that it is impossible for the supercurrent to flow along such a path. Therefore, one can deduce that the magnetic fluxes are arranged at the annulus wall rather than the hollow center of the nano-ring, and form the flux necklace rather than the flux bundle configuration. Note that the vortex cores bounded at the centers of flux quanta is theoretically expected to carry Majorana zero modes due to the topologically nontrivial nature of 2M-$WS_2$, thus providing a practical material platform for understanding vortex control and vortex braiding with interferometric nano-ring structures.

**TDGL simulation of flux distribution**

To understand how the flux configuration evolves with the relation between $\xi_{ab}$ and $w$, we performed theoretical simulations for nanoscale superconducting rings based on the TDGL equation to understand the superconducting order-parameter distribution for steady states (details in Supplementary Fig.

11)[20,29,30]. Figure 3a shows the simulation results of normalized Cooper pair density $|\Psi|^2$ in the nano-ring plane with $w = 4.5\xi_{ab}$, and one can see that the fluxes indeed locate at the center of the nano-ring and do not penetrate through the annulus wall due to the diamagnetism of the superconducting area of the nano-ring (Fig. 3b), consistent with our observations of conventional Little–Parks oscillation. In contrast, when the ratio $w$ to $\xi_{ab}$ increases to $w = 5\xi_{ab}$, the flux quanta (red areas) emerge at the annulus wall of the nano-ring with the radius of $r_{\text{flux}}$ (Fig. 3c), and a maximum of magnetic field distribution appears at the center of annulus wall in the cross-section plot of position-dependent $|\Psi|^2$ and magnetic field profile (Fig. 3d), indicating the formation of flux necklace configuration. The above observations of $\xi_{ab}$-dependent flux bundle-to-necklace transition well explain our experimental observations in Fig. 2b, since cooling the temperature can change the relation between $\xi_{ab}$ and $w$ to switch the flux configuration.

To further understand the evolution of flux distribution as a function of external magnetic field, we analyzed the position-dependent $|\Psi|^2$ value across the annulus wall at different magnetic fields. One can see from Fig. 3e that the $|\Psi|^2$ at the center of annulus wall shows local maxima with the magnetic field up to $0.4H_c$, and switches to local minima with magnetic field above $0.5H_c$, indicating that the superconducting annulus wall starts to accommodate flux quantum with large magnetic field above $0.5H_c$. Such a magnetic field-induced $|\Psi|^2$ variation from local maxima to minima demonstrates that the magnetic field can serve as another degree of freedom to drive the flux configuration transition. This transition can be explained by the distance-dependent repulsive interaction among flux quanta in the following way[27,31]. As magnetic field increases from zero, the first flux quantum will enter the hollow center of the nano-ring. When the magnetic field keeps increasing, multiple flux quanta stay close to each other at the hollow center and the formed flux bundle configuration gradually becomes unstable by considering the free energy of the system. In contrast, with magnetic field further increases above the critical value, the average distance between neighboring flux quanta in flux necklace configuration becomes larger than that of the flux bundle configuration, and the flux necklace configuration with discretely distributed fluxes can have a lower free energy. With the magnetic field further increasing, both the number of flux quanta located at the annulus wall and the size of each flux quantum $r_{\text{flux}}$ gradually increase (Fig. 3f,g), and the flux quanta will eventually contact each other with $H = H_{c2}$ at which the macroscopic phase transition from superconducting to the metallic state appears. Figure 3h shows the magnetic-field-dependent flux quanta number located on the annulus wall at several representative geometrical parameters, and one can see a linear relation between flux number and magnetic field at high magnetic field close to $H_{c2}$, reflecting the type-II superconducting nature of 2M-$WS_2$ nanoflakes.

Figure 3i and Supplementary Figs. 12-14 show the simulated equilibrium states of flux distribution near the flux configuration transition with different geometrical $w/\xi_{ab}$ ratios and magnetic fields. Three important things about such a transition should be addressed here. First, there is a critical value of the inner nano-ring radius $r_{\text{in}}$ for the flux configuration transition. One can see that when the $r_{\text{in}}$ value increases from $5.5\xi_{ab}$ to $5.6\xi_{ab}$ at the fixed magnetic field of $0.4H_c$, the transition from flux bundle to necklace appears (Supplementary Fig. 12), which is in qualitative agreement with our temperature-dependent magnetoresistance evolution results shown in Fig. 2. Second, a critical value of applied field $H$ can also induce such a flux configuration transition. By increasing the strength of magnetic field from $0.453H_c$ to $0.455H_c$ for the fixed geometrical $w/\xi_{ab}$ ratio of 5 (Fig. 3i), a sudden transition to the flux necklace configuration with 5 quantized fluxes simultaneously entering the annulus wall can be observed, in accordance with the simulation results shown in Fig. 3g. Third, although our experimental demonstration is based on circular nano-rings, the flux configuration transition can also be observed in the square ring situation (Supplementary Fig. 15), indicating that this transition is not restricted to the circular geometry but can be generalized to other hollow superconducting nanostructures. This generality highlights the universality of the flux necklace configuration and its potential for diverse device architectures.

To explore the flux distribution within the necklace configuration that might host Majorana bound states on the annulus wall of topological superconducting nano-rings, we performed Bogoliubov-de Gennes calculations[32] for the local density of states (LDOS) distribution in superconducting nano-ring. Note that the calculation is based on the spinless *p*+i*p* wave model, which serves as an equivalent model for simulating the topological superconductivity properties of 2M-$WS_2$ (details in Supplementary Note 12). One can see in Fig. 4a that a ring-shaped zero-energy LDOS is tightly confined in the inner and outer rims of the ring but remains extended along the angular direction. Such results indicate that the superconducting nano-ring can exhibit gapless modes, including the zero-energy Majorana edge modes localized on both the inner and outer edges of the ring, when the flux quanta are located at the nano-ring center at low magnetic fields. As the magnetic field increases, the distribution of flux quanta is changed, and fluxes start to move onto the annulus wall to form the flux necklace configuration. Note that, among those vortex bound states trapped by these vortices, zero-energy Majorana bound modes can be found in the vortex cores of such a flux necklace configuration. These newly-formed Majorana bound modes are fully confined in all dimensions, resulting in a much sharper LDOS in each vortex (Fig. 4b). For clarity, Fig. 4c presents a radial linecut of the zero-energy LDOS through a vortex core. One can see from this linecut that a sharp peak is located at the center for the red curve (obtained from Fig. 4b), while no such peak appears in the blue curve (obtained from Fig. 4a), thus supporting the presence of Majorana bound states with topological protection and strong spatial localization in the vortex cores of the flux necklace configuration.

**Criteria for realizing Majorana edge to bound state transition**

To investigate whether such a flux necklace configuration is a universal phenomenon in nano-rings of type-II superconductors, we performed Little–Parks oscillation study on superconducting nano-rings with conventional *s*-wave superconductor Nb and $NbSe_2$. There are two reasons for choosing these two materials for comparison: First, both Nb and $NbSe_2$ share similar $T_c$ and $\xi_{ab}$ values to those of 2M-$WS_2$, and all these three materials show *s*-wave pairing mechanism in their bulk form[33,34], so comparing the oscillation behavior of these three materials can help understanding the role of topological surface state in the formation of the flux necklace configuration. Second, unlike the highly crystallized nature of $WS_2$ and $NbSe_2$ nano-rings (obtained by mechanical exfoliation), the Nb nano-rings have many disorders that can be the potential pinning centers with which one can understand the role of flux pinning on the formation of the flux necklace configuration by comparing their Little–Parks oscillation behavior. One can see in Fig. 4d, Supplementary Figs. 16,17 and Supplementary Table 3 that the Little–Parks oscillation periods of Nb and $NbSe_2$ nano-rings are in good agreement with the theoretical expectation for flux bundle configuration, and no unconventional oscillations are found for the entire temperature range of superconducting transition, indicating the absence of the flux necklace configuration therein.

We notice that the formation of flux necklace configuration highly relies on the Ginzburg-Landau parameter $\kappa$ ($\kappa = \lambda/\xi_{ab}$, where $\lambda$ is the penetration depth of external magnetic field[35]) and flux pinning energy $U_p$. According to our theoretical simulations based on the TDGL equations, the $\kappa$ value can thermodynamically determine whether the flux necklace configuration can form alongside the nano-ring, while the $U_p$ value can kinetically determine whether the positions of flux quanta can be stabilized for realizing unconventional Little–Parks oscillations (details in Supplementary Figs. 18, 19). Therefore, we summarize the $\kappa$ and $U_p$ values for all the measured samples of three kinds of superconducting materials in this study, as well as other reported topological superconductor candidates[36–41] (Fig. 4e, details in Supplementary Fig. 20 and Supplementary Table 4). One can see that 2M-$WS_2$ simultaneously has both large $U_p$ and large $\kappa$ values that lay the foundation for forming the flux necklace configuration. We emphasize that such a flux configuration transition, could in principle be generalized to nano-rings of all type-II superconductors once the above two conditions are satisfied. Specifically, the conditions for large $U_p$ and large $\kappa$ values can be satisfied in topological superconductor candidates Fe(Te,Se), $UTe_2$ and β-$Bi_2Pd$, in which the Majorana bound states hosted by the vortex cores in flux necklace configuration are expected to form in their nano-ring structures, and might be potentially useful for the braiding of Majorana zero modes. In contrast, the $\kappa$ values of Nb nano-rings are not large enough to thermodynamically form the flux necklace configuration, while the $U_p$ values of $NbSe_2$ nano-rings are not sufficient to stabilize the flux distribution for the interference of Cooper pairs. Therefore, neither Nb nor $NbSe_2$ nano-rings can realize the flux necklace configuration and unconventional Little–Parks oscillation behavior.

Figure 4f and Supplementary Table 5 compare the Little–Parks oscillation and corresponding flux quantization behavior in superconducting nano-rings. Normally, the Little–Parks oscillation in most superconducting rings shows the conventional $\Phi_0 = h/2e$ oscillation period (bold black dashed line), consistent with the formation of Cooper pair carrying $2e$ electrical charges[8–11,24,25,42–55]. However, the oscillation periods of several unconventional superconductors can deviate significantly from $\Phi_0$, falling either below or above the $h/2e$ line. Periods below $h/2e$ are typically attributed to fractionalized flux quantization[2,5] that might arises from multi-charge condensates, such as $4e$- or $6e$-charge condensation[23,56]; in contrast, periods above $h/2e$ are commonly explained by the formation of multi-vortex or giant vortex states, where multiple flux quanta collectively pass through the ring and effectively increase the flux number per period[57–60]. Note that these explanations are established based on the pre-assumption that vortices distribute only in the center of nano-rings, while our observations of flux necklace configuration in 2M-$WS_2$ nano-rings (highlighted by the red circles) demonstrate a new paradigm of flux distribution that can also phenomenologically cause the unconventional Little–Parks oscillation with enlarged oscillation periods. Such a flux necklace configuration in topological superconducting nano-rings is predicted to hold multiple Majorana bound states with potentials for flux braiding towards topological quantum computing.

**Conclusions**

In summary, we demonstrated the realization of flux configuration transition in nano-ring of 2M-$WS_2$

topological superconductor. The observed enlargement of Little–Parks oscillation period corresponds to multiple fluxes distributing along the annulus superconducting region, indicating the emergence of a flux necklace configuration. The Bogoliubov-de Gennes calculations of local density of states further predict that those flux quanta located at the annulus wall of a topological superconducting nano-ring can hold Majorana bound states in their vortex cores. The TDGL simulations reveal that both large flux pinning energy and Ginzburg-Landau parameter are crucial to form and stabilize these vortex bound states within the nano-ring. Note that the switchable flux configuration transition in topological superconductors can serve as a sensitive flux interferometer for reading out Majorana parity towards topological qubit measurement, and the discrete Majorana bound states arranged along the annulus wall provide a natural platform for adiabatic braiding of non-Abelian anyons, a key operation for realizing topological quantum gates of fault-tolerant qubits. Our observations of flux configuration transition in topological superconducting nano-rings provide a new strategy for understanding Majorana modes manipulation through flux-configuration control to realize parity readout and flux braiding towards topological quantum computing.

## Methods

### TDGL simulations for flux quanta distribution

Our simulations of flux dynamics in two-dimensional superconducting rings are based on the nonlinear TDGL equation in the presence of an external field. Under the condition of an external magnetic field, the equilibrium state of the superconductor is determined by its Gibbs free energy, which can be written as[29]:

$$G_{\mathrm{s}} = \iint g_{\mathrm{s}} \mathrm{d}S =$$
$$\iint \left\{ \alpha|\psi|^2 + \frac{\beta}{2}|\psi|^4 + \frac{1}{2m^*}\left|\left(\frac{\hbar}{\mathrm{i}}\nabla - \frac{e^*}{c}\boldsymbol{A}\right)\psi\right|^2 + \frac{(\nabla\times\boldsymbol{A}-\boldsymbol{H})^2}{8\pi} \right\} \mathrm{d}S \tag{3}$$

In this context, $G_{\mathrm{s}}$ denotes the free energy of the superconductor, and $g_{\mathrm{s}}$ represents the superconducting free energy density; $\alpha$ and $\beta$ are phenomenological parameters, which are functions of temperature $T$; $\psi = |\psi|e^{\mathrm{i}\varphi(r)}$ is the superconducting order parameter, and $|\psi|^2 = n_{\mathrm{s}}$ denotes the superconducting electron number density; $c$ is the speed of light in vacuum; $m^*$ and $e^*$ represent the effective mass and electric charge of a Cooper pair, respectively; $\boldsymbol{A}$ is the magnetic vector potential at a certain point on the superconductor, $\boldsymbol{H}$ is the external magnetic induction intensity applied at the same point, and the integration is performed over the entire surface of the superconductor.

By performing variation on $G_{\mathrm{s}}$, the TDGL equation and boundary conditions can be derived.

$$\frac{\hbar^2}{2m^*D}\left(\frac{\partial}{\partial t} + \mathrm{i}\frac{e^*}{\hbar}\varphi\right)\psi = -\frac{1}{2m^*}\left(\frac{\hbar}{\mathrm{i}}\nabla - \frac{e^*}{c}\boldsymbol{A}\right)^2\psi - \alpha\psi - \beta|\psi|^2\psi \tag{4}$$

$$\nabla\times\nabla\times\boldsymbol{A} = -\frac{4\pi\sigma}{c}\left(\frac{1}{c}\frac{\partial\boldsymbol{A}}{\partial t} + \nabla\varphi\right) + \frac{4\pi}{c}\boldsymbol{j}_{\mathrm{s}} \tag{5}$$

$$\boldsymbol{n}\cdot\left(\frac{\hbar}{\mathrm{i}}\nabla-\frac{e^*}{c}\boldsymbol{A}\right)\psi|_{\text{boundary}}=0 \tag{6}$$

$$\boldsymbol{n}\times(\nabla\times\boldsymbol{A}-\boldsymbol{H})|_{\text{boundary}}=0 \tag{7}$$

Among them, $D$ is the diffusion coefficient, $\varphi$ is the scalar potential, $\sigma$ is the material conductivity in the non-superconducting state, and $\boldsymbol{j}_{\mathrm{s}}$ is the supercurrent density:

$$\boldsymbol{j}_{\mathrm{s}}=\frac{\hbar \mathrm{e}^*}{2\mathrm{i}m^*}(\psi^*\nabla\psi-\psi\nabla\psi^*)-\frac{e^{*2}}{m^*c}|\psi|^2\boldsymbol{A} \tag{8}$$

Set $c=1$ and perform dimensionless processing on the variables; finally, the dimensionless TDGL equation and free energy expression[30] can be obtained:

$$\left(\frac{\partial}{\partial t}+\mathrm{i}\varphi\right)\psi=(\nabla-\mathrm{i}\boldsymbol{A})^2\psi+\psi-|\psi|^2\psi \tag{9}$$

$$\kappa^2\nabla\times\nabla\times\boldsymbol{A}=\mathrm{Im}[\psi^*(\nabla-\mathrm{i}\boldsymbol{A})\psi]-\sigma\frac{\partial\boldsymbol{A}}{\partial t} \tag{10}$$

$$G_{\mathrm{s}}=\iint\left\{-|\psi|^2+\frac{1}{2}|\psi|^4+|(\nabla-i\boldsymbol{A})\psi|^2+\kappa^2(\nabla\times\boldsymbol{A}-\boldsymbol{H})^2\right\}\mathrm{d}S \tag{11}$$

Among them, the coordinates are in units of the in-plane coherence length $\xi_{ab}=\hbar/\sqrt{-2m^*\alpha}$; the magnetic field penetration depth is $\lambda_{ab}=\sqrt{-m^*\beta/4\pi e^{*2}\alpha}$; the applied magnetic induction intensity $\boldsymbol{H}$ is in units of the upper critical magnetic field $|\boldsymbol{H}_{\mathrm{c2}}|$ of the superconductor; time is in units of $\tau=\xi_{ab}^2/D$.

Through numerical calculations, the flux necklace configuration emerges at lower temperatures, mainly due to the competition between superconducting condensation energy $g_{\mathrm{sup}}=-|\psi|^2+\frac{1}{2}|\psi|^4$ and interaction energy $g_{\mathrm{int}}=|(\nabla-i\boldsymbol{A})\psi|^2$. At low temperatures ($w=5\xi_{ab}$), the $g_{\mathrm{int}}$ increase from quantized magnetic fluxes entering the superconductor is smaller than the $g_{\mathrm{sup}}$ decrease from superconducting state destruction at the flux center; thus, fluxes tend to enter the superconductor in a quantized form and distribute into the minimum $g_{\mathrm{int}}$ state, the flux necklace configuration.

It should be noted that in the flux dynamics simulations based on the TDGL equation, the phenomenon of flux rearrangement may sometimes occur, and the change of flux position with time is not obvious. Therefore, it is necessary to define a quantitative criterion for the system to reach a steady state. For the simulation under a constant magnetic field, the criterion for the system to reach a steady state is that the change of $G_{\mathrm{s}}$ is less than 0.1% within a time of $100\tau$; for the simulation under a sweeping magnetic field, due to the small magnetic field step size ($\Delta H = 0.001H_{\mathrm{c2}}$) , the change of the system state is small, so maintaining each magnetic field for a time of $400\tau$ is sufficient for the system to reach the corresponding steady state.

The pinning potential energy $U_{\mathrm{p}}$, as a critical criterion for the existence of the flux necklace state, is obtained by fitting experimental data with the Thermally Activated Flux Flow model:

$$R_{xx}(T) = R_0 \exp\left(-\frac{U_{\mathrm{p}}(T)}{k_{\mathrm{B}}T}\right) \quad (12)$$

where $k_{\mathrm{B}} = 8.617 \times 10^{-5}\ \mathrm{eV/K}$ is the Boltzmann constant and $R_0$, $U_{\mathrm{p}}$ are fitting parameters. Typically, $U_{\mathrm{p}}$ represents the thermal activation energy required for a flux quantum to escape the pinning potential well. Meanwhile, in the TDGL simulations incorporating flux pinning, the corresponding form of the pinning potential can be given by:

$$U_{\mathrm{p}} = \iint \{g_{\mathrm{np}}(\psi_{\mathrm{np}}, A_{\mathrm{np}}, \gamma = 0) - g_{\mathrm{p}}(\psi_{\mathrm{p}}, A_{\mathrm{p}}, \gamma = \gamma(\boldsymbol{r}))\}\, \mathrm{d}S \quad (13)$$

where $g$ denotes the energy density and $\gamma(\boldsymbol{r})$ is a dimensionless pinning parameter ranging from 0

(no pinning) to 1 (very strong pinning). Detailed discussions of $U_p$ and related simulations are shown in Supplementary Notes 15 and 16.

**Bogoliubov-de Gennes simulations for local density of states**

Bogoliubov-de Gennes calculations are applied to investigate the real-space distribution of local density of states zero-energy modes in a mesoscopic superconducting ring with both flux bundle and necklace configurations. As an equivalent model to simulate the topological superconducting nature of 2M-$WS_2$, we considered a spinless $p$+i$p$ wave superconductor described by the continuum Hamiltonian

$$H = \int \mathrm{d}^2\vec{r}\, \psi^+ \left(-\frac{\nabla^2}{2m} - \mu\right)\psi + \left[\psi^+ \Delta(\vec{r})\left(-i\partial_x + \partial_y\right)\psi^+(\vec{r}) + h.c.\right], \tag{14}$$

where the pairing field $\Delta(\vec{r})$ is determined by the vortex configuration

$$\Delta(\vec{r}) = \Delta_0 \prod_i \tanh\frac{\left|\vec{r} - \vec{R}_i\right|}{\xi} e^{-i\arg(\vec{r}-\vec{R}_i)} \tag{15}$$

($\vec{R}_i$ is the center of the $i$-th vortex and $\xi$ is the superconducting coherence length). After discretizing Eq. (14) up to the nearest neighbors and working in Nambu space, the Bogoliubov-de Gennes Hamiltonian can be written as $\boldsymbol{H} = \Psi^+ \boldsymbol{h} \Psi$, where $\boldsymbol{h}$ is a $d \times d$ Hermitian matrix with $d = N_r \times N_\theta \times 2$. We finally calculate the local density of states at zero energy

$$\rho_{ij}(0) = \frac{1}{\pi}\sum_n \frac{|u_n(ij)|^2\eta}{E_n^2 + \eta^2}, \tag{16}$$

where $\eta$ is a smearing factor accounting for energy level broadening due to environmental fluctuation or disorders.

**Fabrication and transport measurements of interferometric superconducting nano-rings**

2M-$WS_2$ nanoflakes with desired thickness were prepared via mechanical exfoliation in a $N_2$ glove box (oxygen and water <0.1 ppm) and subsequently transferred onto Si/$SiO_2$ substrates using a polydimethylsiloxane stamp. Polymethyl methacrylate was then spin-coated on the substrate and put in a vacuum drying oven kept at <0.01 MPa without heating the sample, which can remove the solvent in polymethyl methacrylate but avoid introducing any phase transition in 2M-$WS_2$ nanoflakes. A four-terminal electrode configuration was fabricated by electron beam lithography (EBL), followed by the deposition of Ti (5 nm) and Au (60 nm) electrodes by metal evaporator. A second electron beam lithography step was employed to define single-ring patterns on the nanoflakes. The devices were then processed by reactive ion etching for 3 min ($CF_4$ flow: 30 sccm; pressure: 4 Pa) with a power of 30 W to form the superconducting nano-rings.

The magnetoresistance oscillation of the single-ring devices was measured using an alternating current method with SR830 lock-in amplifiers (Stanford Research Systems) at the frequency of $f = 13$ Hz. The $I$-$V$ characteristics of the devices and $I_c$ evolutions were studied by the direct current method with a sourcemeter (Keithley 2400) and a nanovoltmeter (Keithley 2182A). All measurements were carried out in an Oxford Instruments Teslatron$^{PT}$ capable of reaching the lowest temperature of 1.5 K and applying magnetic fields up to 12 T. Note that the fluctuation of temperature is controlled within the range of ±1 mK during magnetoresistance measurements to minimize the thermal noise and capture the details of the magnetoresistance signals.

**Data availability**

All data needed to evaluate the conclusions in the paper are presented in the paper or the supplementary materials and are available from the corresponding author upon reasonable request.

## Acknowledgements

The authors would like to acknowledge the support by the National Natural Science Foundation of China (Nos. 92365203 (H.T.Y.), U24A6002 (H.T.Y.), 52302180 (J.W.H.), 12374147 (Q.H.W.), 12274205 (D.W.) and 92565109 (Y.Q.F.)), the National Key Research and Development Program of China (Nos. 2024YFA1408104 (H.T.Y.), 2021YFA1202901 (J.W.H.), 2022YFA1403201 (D.W.) and 2025YFF0516300 (F. Q. H)), the Quantum Science and Technology-National Science and Technology Major Project (No. 2021ZD0302502 (H.T.Y.)), the Natural Science Foundation of Jiangsu Province (Nos. BK20253012 (H.T.Y.), BK20233001 (H.T.Y.) and BK20243011 (H.T.Y.)). the Guangdong Provincial Quantum Science Strategic Initiative (GDZX2501002 (Z.L.), the Guangdong Basic and Applied Basic Research Foundation (2026A1515011759 (Z.L.)), the Shenzhen Municipal Funding Co-Construction Program Project (SZZX2501001 (Z.L.)), the Science and Technology Commission of Shanghai Municipality (No. 24LZ1401000 (Y.Q.F.)).

## Author Contributions

H.T.Y., Y.I. and Q.H.W. conceived the project. X.Y.B. and A.S.Y. prepared the superconducting nano-ring devices. X.Y.B., A.S.Y. and J.W.H. performed low-temperature electronic transport measurements. Z.Y.L. and Q.Z.S. performed AFM measurements. Y.Q.F. and F.Q.H. prepared the high-quality 2M-$WS_2$ crystal. H.Z.C. and G.Y.C. presented theoretical modeling of flux states in superconducting nano-rings. D.W. and Q.H.W. performed Bogoliubov-de Gennes calculations. X.Y.B., A.S.Y., H.Z.C., T.I. and H.T.Y. analyzed the data. X.Y.B., A.S.Y. and H.T.Y. wrote the manuscript with inputs from all authors.

## Competing interests

The authors declare no competing interests.

**Correspondence and requests for materials** should be addressed to H.T.Y..

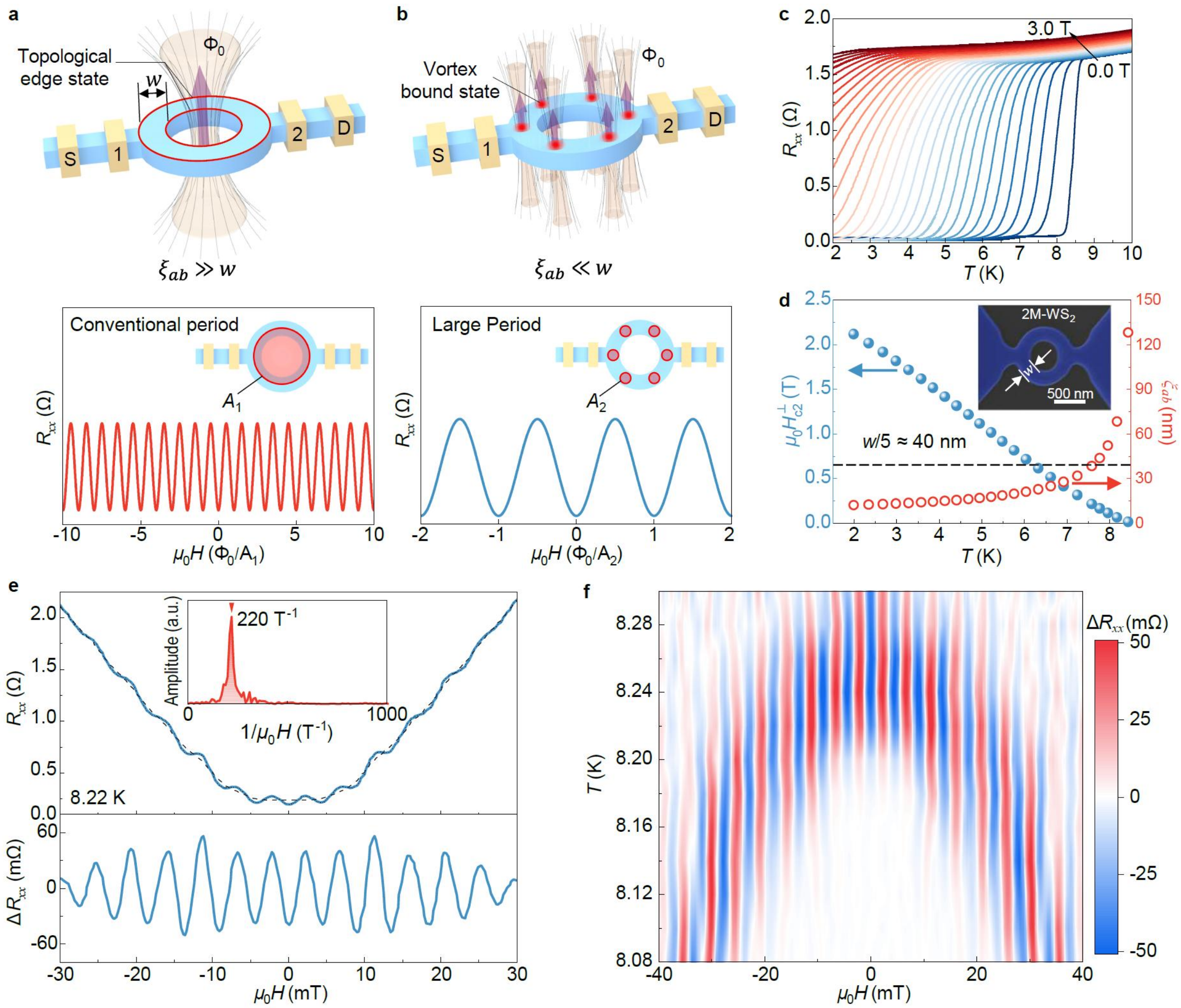

**Fig. 1 | Two flux configurations with different topological states in 2M-$WS_2$ nano-rings. a**, Schematic illustration of the flux bundle configuration and the conventional Little–Parks oscillation. The flux quantum $\Phi_0$ (light purple arrows) is arranged at the center of the nano-ring. Electrodes S and D are source and drain of electrical current, and 1, 2 are voltage probes. The effective supercurrent interference area is marked as $A_1$ in the inset of the lower panel. **b**, Schematic illustration of the flux necklace configuration and the unconventional Little–Parks oscillation with enlarged period. The effective supercurrent interference area is marked as $A_2$. **c**, Temperature-dependent $R_{xx}$ of a typical nano-ring device at different out-of-plane magnetic fields. **d**, Temperature dependence of the upper

critical magnetic field $\mu_0 H_{c2}^{\perp}$ (blue ball) and the in-plane coherence length $\xi_{ab}$ (red circle). The black dashed line highlights the condition of $\xi_{ab}$ = *w*/5. Inset: False-colored scanning electron microscopy image of the fabricated 2M-$WS_2$ nano-ring device. The scale bar is 500 nm. **e**, Upper panel: Magnetoresistance of the nano-ring device, where the background signal can be subtracted by the Savitzky-Golay fitting (black dashed line). Lower panel: Pure oscillation signal $\Delta R_{xx}$ after subtracting the background. Inset: Fourier transform of $\Delta R_{xx}$. **f**, Mapping results of $\Delta R_{xx}$ for the flux bundle configuration at different temperatures.

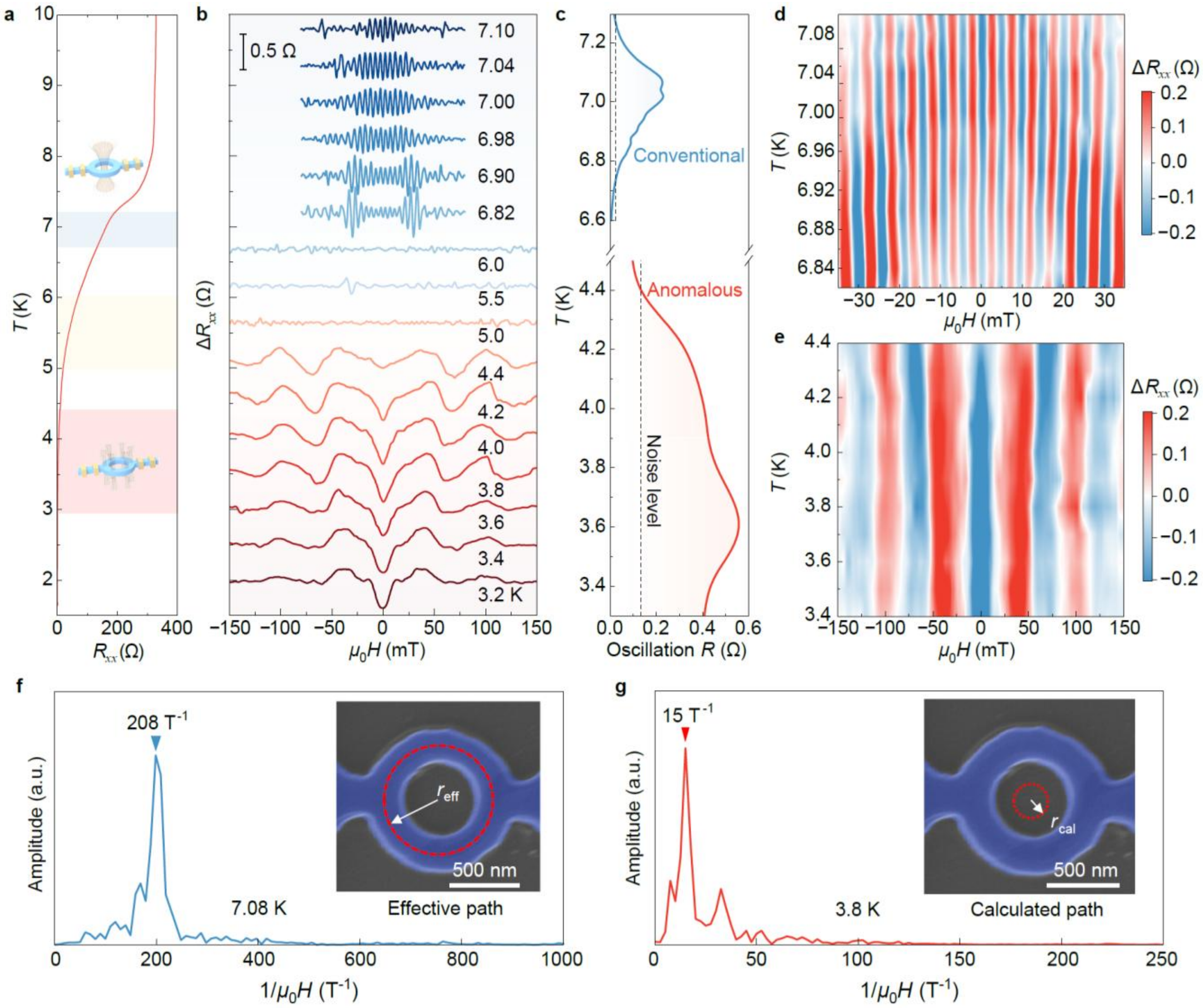

**Fig. 2 | Flux bundle to necklace configuration in thin 2M-$WS_2$ nano-rings. a**, Temperature dependence of $R_{xx}$. Light blue and red regions correspond to the temperature ranges that convention and unconventional oscillation phenomena were observed, and the light yellow region corresponds to the temperature range that no apparent oscillation signal can be found. **b**, $\Delta R_{xx}$ at different temperatures after subtracting the background magnetoresistance signal. The temperature range of 6.82~7.10 K (3.2~4.4 K) corresponds to the conventional flux bundle (unconventional flux necklace) configuration, and the temperature range of 5.0~6.0 K shows no apparent oscillation signal. **c**, Comparison of the oscillation amplitude for the conventional (blue) and unconventional (red) Little–Parks oscillations. **d**,**e**, Mapping results of conventional and unconventional Little–Parks oscillations

at different temperatures. **f**,**g**, Fourier transform results for the conventional and unconventional Little–Parks oscillation. The calculated effective coherence paths (red dashed circles) for both types of oscillations are sketched in the inset scanning electron microscopy figures. The scale bars are 500 nm.

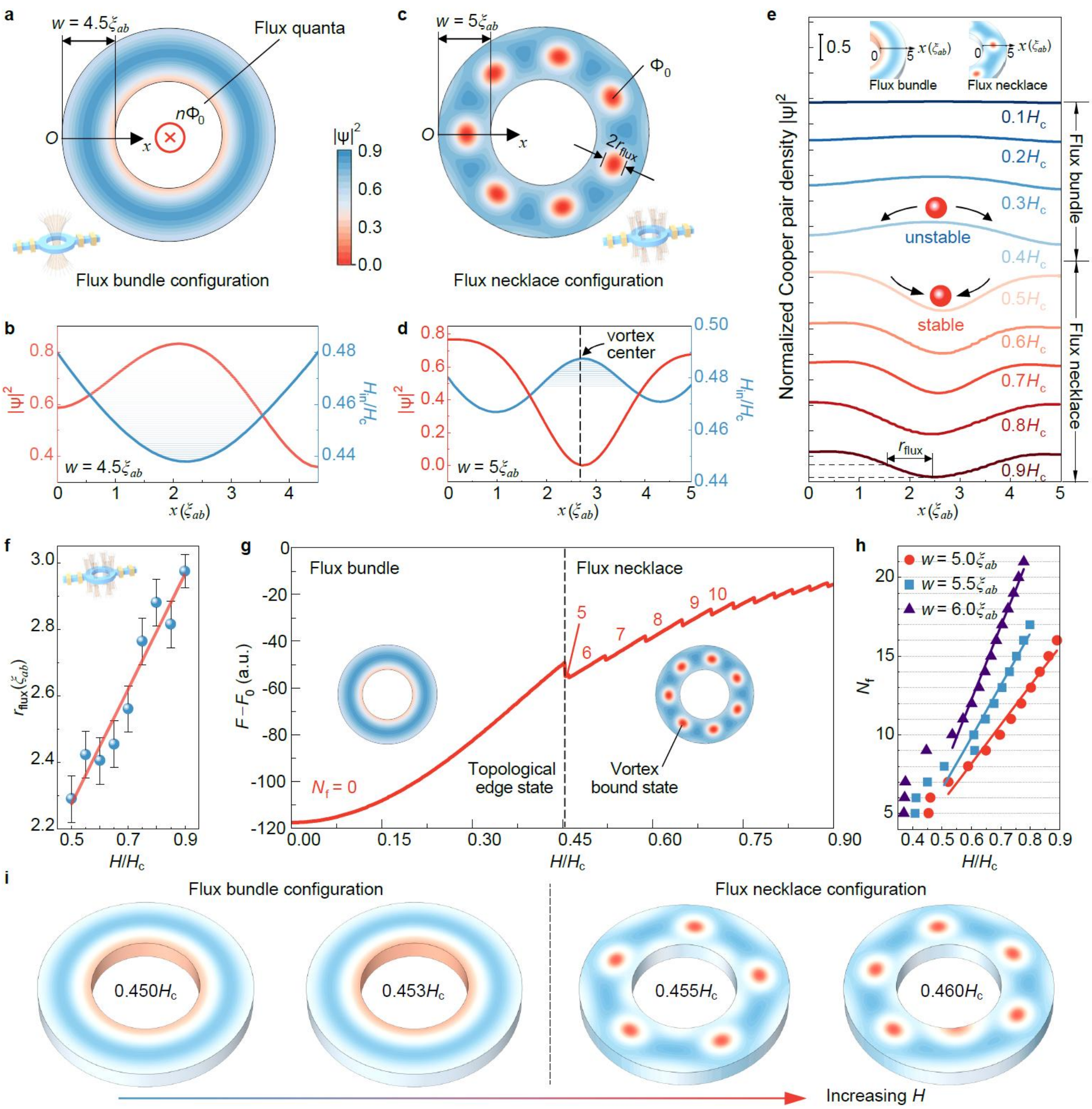

**Fig. 3 | TDGL simulations of flux distribution in a nano-ring. a**, Simulation results for the flux bundle configuration when $w$ = $4.5\xi_{ab}$ and $H$ = $0.48H_c$. The color bar represents the value of the superconducting order parameter $|\Psi|^2$. **b**, Distribution of $|\Psi|^2$ and local magnetic field in the annulus wall along the $x$ axis in (a). **c**, Simulation results for the flux necklace configuration when $w$ = $5.0\xi_{ab}$ at the same condition of $H = 0.48H_c$. **d**, Distribution of $|\Psi|^2$ and local magnetic field in the annulus wall crossing the flux center in (c). **e**, Local distribution of $|\Psi|^2$ along the annulus wall under

increasing magnetic fields for $w = 5\xi_{ab}$. Inset panels schematically demonstrate the line cut for both flux bundle (left panel) and flux necklace (right panel) configurations. **f**, Simulated magnetic-field-dependent flux core radius $r_{\text{flux}}$ at the fixed geometric relation of $w = 5\xi_{ab}$. **g**, Simulated magnetic-field-dependent number of flux quanta $N_f$ on the annulus wall. An abrupt transition from 0 (flux bundle, with only conventional fluxes) to 5 (flux necklace, with fluxes carrying Majorana zero modes) can be observed when $H > 0.45H_c$. **h**, Simulated magnetic field dependence of quantum flux filling number $N_f$ at different geometric conditions. **i**, Simulated phase transition from flux bundle to flux necklace configuration as $H$ increases in (g).

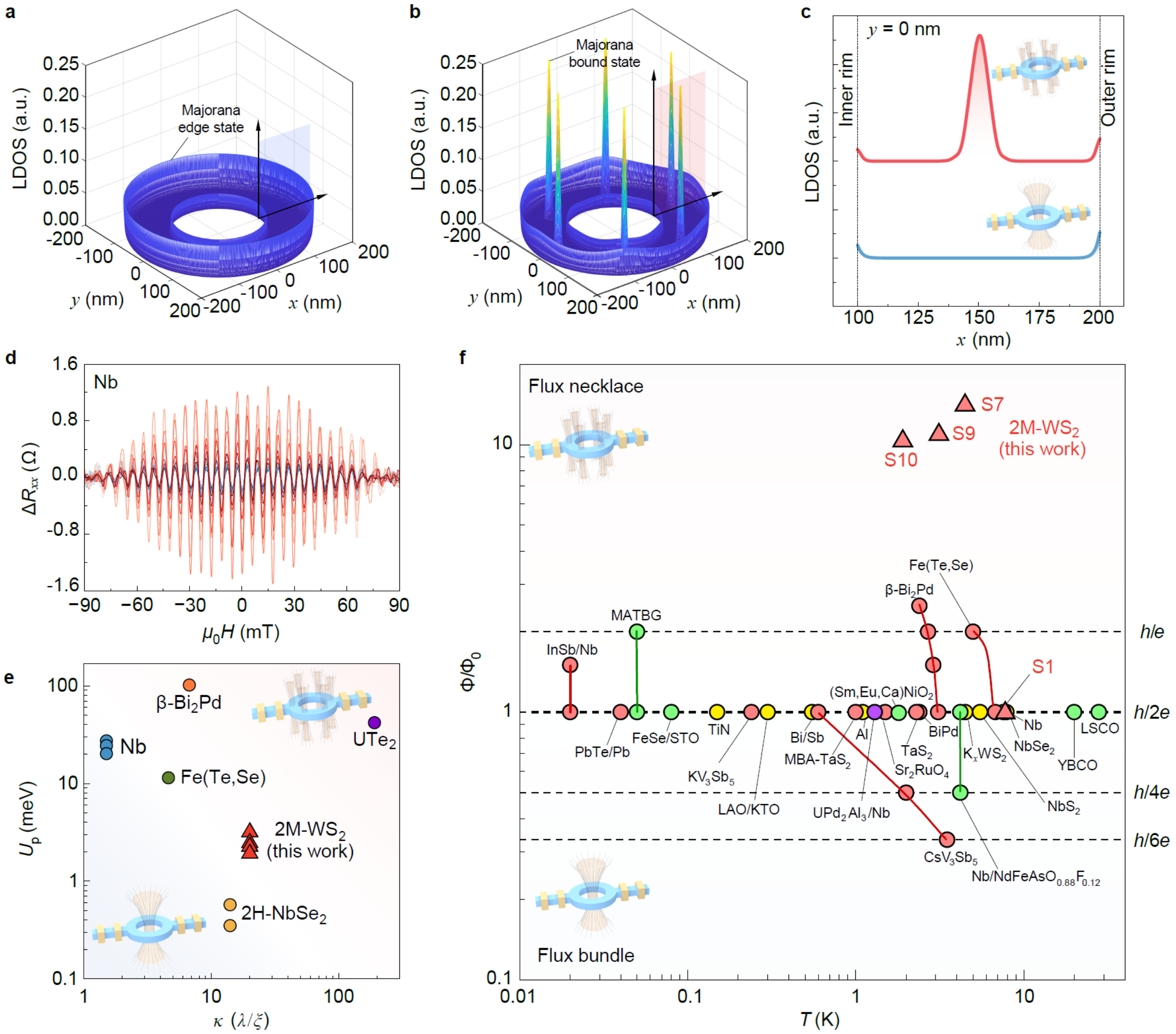

**Fig. 4 | Comparison of magnetic flux distribution in emergent superconducting nanostructures.** **a**, Calculated real-space distribution of LDOS of a *p*-wave superconducting nano-ring for the flux bundle configuration. Both the inner and outer rims can accommodate an Majorana edge state. **b**, Calculated LDOS of a *p*-wave superconducting nano-ring for the flux necklace configuration. The cusp peaks represent the Majorana bound states. **c**, Radial linecut of LDOS distribution in (a) (blue curve, corresponding to light blue cross-section) and (b) (red curve, corresponding to light red cross-section). **d**, Conventional Little–Parks oscillation of a Nb nano-ring device at different temperatures in its superconducting transition regime. **e**, Comparison among our experimental results for $WS_2$,

$NbSe_2$ and Nb nano-rings, as well as other topological superconductor candidates. **f**, Summary of the oscillation periods and their observed temperatures for our 2M-$WS_2$ nano-rings (red triangles) and other reported systems (circles). Yellow circles: *s*-wave superconductors; green circles: high-$T_c$ superconductors; red circles: topologically nontrivial superconductors; purple circles: heavy-fermion superconductors. The oscillation period significantly larger than expectation of $h/2e$ might correspond to the giant vortex or the flux necklace configuration.